\newcommand{\slos}{\sigma_{\rm LOS}}

\documentclass[apj,iop]{emulateapj}

\slugcomment{11 August 2011, accepted to ApJ Letters}

\begin{document}

\title{Galaxy Disks are Submaximal}

\author{
Matthew A. Bershady,\altaffilmark{1}
Thomas P. K. Martinsson,\altaffilmark{2}
Marc A. W. Verheijen,\altaffilmark{2}
Kyle B. Westfall,\altaffilmark{2,3}
David R. Andersen,\altaffilmark{4} and
Rob A. Swaters\altaffilmark{5}
}

\altaffiltext{1}{University of Wisconsin, Department of Astronomy, 475
N. Charter St., Madison, WI 53706; mab@astro.wisc.edu}

\altaffiltext{2}{University of Groningen, Kapteyn Astronomical
  Institute, Landleven 12, 9747 AD Groningen, Netherlands}

\altaffiltext{3}{International Research Fellow, National Science Foundation}

\altaffiltext{4}{NRC Herzberg Institute of Astrophysics, 5071 W
Saanich Road, Victoria, BC V9E 2E7, Canada}

\altaffiltext{5}{National Optical Astronomical Observatories,
Tucson, AZ}

\begin{abstract}
  We measure the contribution of galaxy disks to the overall
  gravitational potential of 30 nearly face-on
  intermediate-to-late-type spirals from the DiskMass Survey. The
  central vertical velocity dispersion of the disk stars
  ($\sigma_{z,R=0}^{\rm disk}$) is related to the maximum rotation
  speed ($V_{\rm max}$) as $\sigma_{z,R=0}^{\rm disk} \sim 0.26 V_{\rm
    max}$, consistent with previous measurements for edge-on disk
  galaxies and a mean stellar velocity ellipsoid axial ratio $\alpha
  \equiv \sigma_z / \sigma_R = 0.6$.  For reasonable values of disk
  oblateness, this relation implies these galaxy disks are
  submaximal. We find disks in our sample contribute only 15\% to 30\%
  of the dynamical mass within 2.2 disk scale-lengths ($h_R$), with
  percentages increasing systematically with luminosity, rotation speed and
  redder color. These trends indicate the mass ratio of disk-to-total
  matter remains at or below 50\% at 2.2 $h_R$ even for the most
  extreme, fast-rotating disks ($V_{\rm max}\geq 300$ km s$^{-1}$), of
  the reddest rest-frame, face-on color ($B-K \sim 4$ mag), and
  highest luminosity ($M_K<-26.5$ mag). Therefore, spiral disks in
  general should be submaximal. Our results imply that the stellar
  mass-to-light ratio and hence the accounting of baryons in stars
  should be lowered by at least a factor of 3.
\end{abstract}


\section{INTRODUCTION}

Rotation-curve decomposition is the primary tool for measuring the
distribution of dark matter in spiral galaxy halos, but the tool is
blunted by uncertainties in the mass-to-light ratio ($\Upsilon$) of
the luminous disk and bulge. The `maximum-disk hypothesis' (van Albada
\& Sancisi 1986; see also Binney \& Tremaine 2008) bypasses this
mass-decomposition (`disk-halo') degeneracy. A disk contributing
maximally to the gravitational potential sets a lower limit on the
amount of halo dark matter in the inner regions of disk
galaxies. Maximum-disk decompositions find the disk mass produces
$85\pm10$\% of the observed rotation velocity at 2.2 disk
scale-lengths ($h_R$; Sackett 1997). Unfortunately, this hypothesis
remains unproven, and there have been suggestions to the contrary
based on the lack of surface-brightness dependence in the Tully-Fisher
relation (TF; Tully \& Fisher 1977) for a wide range of spirals
(Courteau \& Rix 1999; Courteau et al. 2003; Zwaan et
al. 1995). Constraints on $\Upsilon$ from stellar population synthesis
(SPS) models are poor due to long-standing uncertainties in the
low-mass end of the stellar IMF and late phases of stellar evolution
(e.g., TP-AGB stars; Maraston 2005, Conroy et al. 2009).

Direct kinematic evidence for the mass contribution from spiral disks
stems from a series of longslit spectroscopic studies (van der Kruit
\& Freeman 1984; Bottema 1993; Kregel et al. 2005) showing that the
ratio of the disk stellar velocity dispersion and the maximum observed
rotation speed was much lower than expected for a maximum
disk. Bottema (1993) found disks contribute only $63\pm10$\% of the
observed rotation speed. Most of these galaxies are viewed edge-on,
however, so that the inference of disk mass requires some uncertain
assumptions about the stellar velocity ellipsoid (SVE) and
line-of-sight deprojection. Recent results from PNe kinematics of
several nearby galaxies suggest disk-maximality may depend on Hubble
type (Herrmann \& Ciardullo 2009).  Constraints on disk maximality
from hydrodynamical modeling of observed non-axisymmetric gas flows
has conflicted (cf. Weiner et al. 2001 and Kranz et al. 2003).
Analyses of one gravitional-lens system (Dutton et al. 2011) and one
barred, resonance-ring system (Byrd et al. 2006) indicate the disks in
these two galaxies are submaximal. Observational evidence suggests
disks are submaximal, but the question is unsettled.

The DiskMass Survey (DMS; Bershady et al. 2010a) breaks the disk-halo
degeneracy by obtaining independent measures of the total dynamical
mass and dynamical disk-mass surface density from integral-field
stellar and gas kinematics of face-on, disk-dominated
galaxies. Following the approach of Bottema (1993) and Kregel et
al. (2005), in this letter we derive the disk maximality based on the
relation between the disk vertical stellar velocity dispersion
($\sigma_z$) and the disk-equatorial circular speed of the potential,
making basic assumptions of disk equilibrium and a testable hypothesis
about the SVE. We adopt $H_0 = 73$ km s$^{-1}$ Mpc$^{-1}$ and Vega
magnitudes.

\section{Data and Measurements}

For 30 nearly face-on spiral galaxies covering a range in morphology,
mass, color, surface brightness and scale length from our Phase-B
sample (Bershady et al. 2010a), the stellar line-of-sight velocity
dispersions $\slos$ and gas rotation curves were measured out to $\sim
3h_R$. Spectroscopic data in the MgI region (498-538 nm) at
resolutions of $\lambda/\Delta\lambda \sim 7700$ were collected with
the PPak integral field unit (Verheijen et al. 2004; Kelz et al. 2006)
of the PMAS spectrograph (Roth et al. 2005) on the 3.5m Calar Alto
telescope.\footnote{Centro Astron\'omico Hispano Alem\'an (CAHA) at
  Calar Alto, operated jointly by the MPIA and CSIC.}
Stellar velocity dispersions were measured down to $\slos
\sim 18$ km s$^{-1}$ for individual fiber spectra via our
data-censored cross-correlation software (DC3; Westfall et al. 2011a)
using a K1~III stellar template.

The maximum rotation speed of the potential ($V_{\rm max}$) at
$2.2h_R$ is derived by fitting a simple hyperbolic-tangent (tanh)
model (e.g., Andersen et al. 2001) to the two-dimensional
[OIII]$\lambda 5007$ velocity field, deprojected using inclinations
($i_{TF}$) inferred from inverting the Tully-Fisher relations from
Verheijen (2001) for the $K$-band. To minimize systematic error, total
magnitudes estimation matched the protocol used by Verheijen (2001),
using elliptical-aperture surface-photometry of reprocessed 2MASS
images (Skrutskie et al. 2006), as described in Westfall et
al. (2011b; DMS-IV). Since $V_{\rm max}$ is derived from the
asymptotic velocity of the tanh model, systematic errors arise if the
rotation curve is not truly flat. Inspection of each galaxy's
position-velocity diagram shows these errors to be of the same order
as the formal errors from the fitting, in most cases. In Figure 1,
below, we flag two exceptions: UGC 1862, which still has a rising
rotation curve at the outermost point; and UGC 4458, which has a
declining rotation curve after reaching $V_{\rm max}$ due to a
prominent bulge. Random errors on $V_{\rm max}$ (7\% on average) are
dominated by inclination uncertainties.

The central value of the stellar velocity dispersion of the {\it disk}
($\sigma_{\rm LOS}^{\rm disk}$) is found by fitting a radial
exponential function to the $\slos$ measurements in a radial range
that shows an exponential decline (typically $0.5<R/h_R<4$), e.g.,
excluding the bulge region, and then extrapolating to $R=0$. The
exponential function is found to fit well in this radial range.
Because $\sigma_z$ is expected to scale with the square root of the
surface density, the kinematic scale-length ($h_\sigma$) is expected
to be $2h_R$ for an exponential disk of constant thickness and
$\Upsilon$ viewed perfectly face-on. In our fitting, we allow
$h_\sigma$ to be a free parameter; the implications for $\Upsilon$
disk gradients are discussed in Martinsson (2011). For present
purposes it suffices to state $\Upsilon$ gradients are generally
small.

The vertical component $\sigma_{z}^{\rm disk}$ of the stellar velocity
ellipsoid is extracted from $\sigma_{\rm LOS}^{\rm disk}$ by assuming
a constant shape of the SVE and $i_{TF}$. Following Bershady et
al. (2010b; DMS-II) we adopt $\alpha = 0.6 \pm 0.15 \ (25\%)$ and
$\beta \equiv \sigma_{\theta} / \sigma_R = 0.7 \pm 0.04 \ (5\%)$, where $\sigma_R$ and
$\sigma_{\theta}$ are the radial and tangential components of the SVE.  These
are reasonable values given extant results for external galaxies
summarized by van der Kruit \& de Grijs (1999), Shapiro et al. (2003),
and our own work (Westfall 2009 and DMS-IV).  We may then write
$\sigma_z^2 = \frac{\slos^2}{\gamma\cos^2i}$; $\gamma$ is a projection
factor defined in DMS-II as a function of the SVE ratios and
inclination. A typical value of $\sqrt{\gamma} \cos i$ is 1.1, with a
10\% uncertainty. This factor dominates the error on $\sigma_z$ at small
radii and larger inclinations, with the largest contribution from the
uncertainty in $\alpha$. Errors contributed by inclination
uncertainties are lower but non-negligible; errors contributed by
$\beta$ computed from the epicycle approximation are negligible.
Measurement error in $\slos$ generally dominate at larger radii and at
lower inclinations.  Details of the sample, instrumental
configurations, data acquisition and reduction, basic data products,
and full mass decompositions are presented in Martinsson (2011).

\begin{figure*}
\centering
\leavevmode
\includegraphics[scale=0.7]{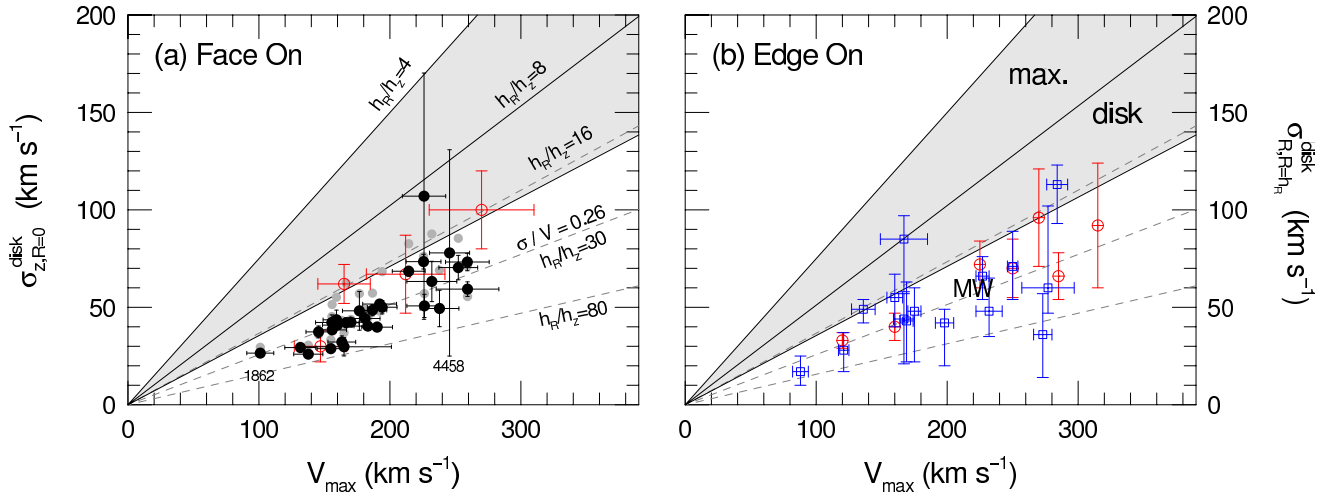}
\caption{Stellar velocity dispersion versus maximum gas
  rotation speed: (a) $\sigma_{z,R=0}^{\rm disk}$ for face-on galaxies
  from the DiskMass survey (filled circles) and Bottema (1993; open
  circles); (b) $\sigma_{R,R=h_R}^{\rm disk}$ for edge-on galaxies
  from Bottema (1993) and Kregel et al. (2005; open squares and MW for
  the Milky Way). Lines show the maximum rotational velocity allowed
  for a self-gravitating galactic disks of different oblateness
  (labeled). Maximum disks in the observed oblateness range would fall
  in the gray shaded region.}
\end{figure*}

\section{Disk Maximality}

For an oblate, self-gravitating, exponential disk of constant
$\Upsilon$ it can be shown (Freeman 1970, Casertano 1983, Kuijken \&
Gilmore 1989, DMS-II) that the maximum rotation speed of the disk is
related to the disk central mass surface-density, $\Sigma_0$, and
$h_R$:
\begin{equation}
V_{\rm max}^{\rm disk} \ = \ c_{\rm max} \sqrt{\pi~G~\Sigma_0~h_R}
\label{eq:VS_flat1}
\end{equation}
where $c_{\rm max} \sim 0.88 (1-0.28/q)$ to excellent approximation
for $q<4$, and $q \equiv h_R/h_z$ is the oblateness parameter.  A
comparison of $V_{\rm max}$ to $V_{\rm max}^{\rm disk}$ is an accurate
measure of {\it disk} maximality regardless of a bulge component. The
comparison also provides a good estimate of baryon maximality because
$V_{\rm max}^{\rm disk}$ occurs at $R/h_R \sim 2.2$ where most bulges
do not significantly contribute to $V_{\rm max}$.

For a disk in equilibrium $\Sigma_{\rm dyn} = \sigma_z^2 / \pi k G
h_z$, where $k$ parameterizes the vertical density distribution with
likely values between 1.5 to 2 (exponential to isothermal
distributions, respectively; van der Kruit 1988).  Because we measure
$\slos$ outside of the galaxy center, where the bulge contribution is
minimal and it is plausible that the SVE is relatively constant with
radius, our estimate of $\sigma_{z,R=0}^{\rm disk}$ is well-posed. In
contrast, central measurements 
of $\slos$ are contaminated by bulges, bars, and non-circular
motions. For a disk embedded in a dark halo, we expect $\sigma_z$ will
increase for a given disk $\Sigma_{\rm dyn}$ (Bottema 1993), hence the
estimate is an upper limit.

Combining Equation 1 with the expression for $\Sigma_{\rm dyn}$, and
normalizing $k$ to an exponential density distribution and $q$ for an
oblateness of a typical disk with a 3.5 kpc scale-length (Equation 1
from DMS-II), we find the following relation between the {\it
  maximum} rotation speed of the disk, {\it disk} central
velocity dispersion, and disk oblateness:
\begin{equation}
V_{\rm max}^{\rm disk}=\left[2.04\left(\frac{q}{8.1}\right)
-0.07\right]\left(\frac{q}{8.1}\frac{k}{1.5}\right)^{-0.5}
\sigma_{z,R=0}^{\rm disk},
\label{eq:VS_flat2}
\end{equation}
Our expectations are, then, that for a maximum disk where $V_{\rm
max}^{\rm disk} = 0.85 V_{\rm max}$, there should be a linear relation
between $V_{\rm max}$ and $\sigma_{z,R=0}^{\rm disk}$ with a slope of
$\sim$0.43. 

However, in Figure 1a, the distribution of $\sigma_{z,R=0}^{\rm disk}$
versus $V_{\rm max}$ for the DMS sample lie near $\sigma_{z,R=0}^{\rm
disk} = (0.26 \pm 0.10) V_{\rm max}$, a slope well below the expected
value of 0.43 for a maximum disk. The disks of our sample galaxies
would be maximal if $h_R/h_z \sim 30$. Based on the compilation from
DMS-II (e.g., see Kregel et al. 2002), this oblateness is twice as flat
as any observed edge-on disk.  For realistic values of $h_R/h_z$, the
disk-only rotation curves peak well below the measured $V_{\rm max}$.
{\it These galaxy disks are sub-maximal.}

To test this result, we plot the uncorrected central disk velocity
dispersion ($\sigma_{LOS,R=0}^{\rm disk}$) as filled gray points in
Figure 1a. The corrections for the SVE projection are relatively
small. Even without the correction disks still appear to lie in a
parameter space that is marginally sub-maximal. This is an upper limit
for any reasonable SVE with $\alpha<1$ and $\beta$ within a factor of
2 of the epicyclic value, and considering the above-mentioned impact
of a dark halo on $\sigma_z$.

In Figure 1b we plot $\sigma_{R,R=h_R}^{\rm disk}$ versus $V_{\rm
max}$ for the edge-on samples of Bottema (1993) and Kregel et
al. (2005). As these authors state, for an exponential disk
mass-density distribution of constant SVE with $\alpha = 0.6$,
$\sigma_{R,R=h_R}^{\rm disk} \sim \sigma_{z,R=0}^{\rm disk}$. The
similarity of the relation found for both edge-on and face-on samples
demonstrates the accuracy of our basic assumption concerning the
average SVE: If $\alpha$ were 65\% larger on average ($\alpha=1$
instead of 0.6), then the edge-on sample would be centered in the
maximum-disk region of Figure 1. While this check is of primary
importance for interpreting the maximality of edge-on disks, it also
confirms our second-order corrections to $\sigma_{LOS}^{\rm disk}$ for
face-on samples. A corollary, important below, is that the
disk-oblateness measurements of these edge-on samples, used for
mass-estimates of our face-on sample, are likely also an accurate
application.

\begin{figure*}
\centering
\leavevmode
\includegraphics[scale=0.75]{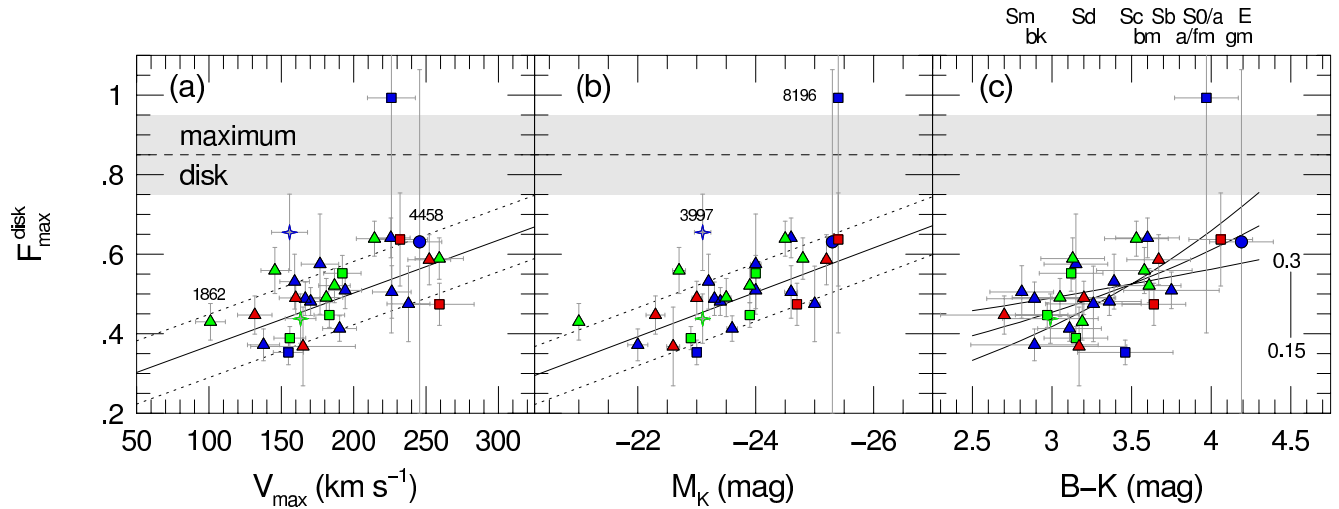}
\caption{Ratio of disk rotation speed to total rotation speed (F$_{\rm
    disk}$) versus (a) total rotation speed, (b) K-band luminosity and
  (c) $B-K$ color for DiskMass Survey galaxies. Maximum disks have
  F$_{\rm disk} = 0.85 \pm 0.1$ (shaded gray). Galaxies are coded by
  Hubble type (Sa, circles; Sb/Sbc, squares; Sc/Scd, triangles; Sd/Im,
  pluses) and bar classification (SA, blue/dark gray; SAB, green/light
  gray; SB, red/medium gray).  Lines and labels are described in the
  text.}
\end{figure*}

\section{Why we expect all disks are submaximal}

We quantify disk-maximality as $F_{\rm max}^{\rm disk} \equiv V_{\rm
  max}^{\rm disk} / V_{\rm max}$, using Equation 2 for $V_{\rm
  max}^{\rm disk}$ and Equation 1 from DMS-II for $q$.  Figure 2 shows
$F_{\rm max}^{\rm disk}$ of the DMS sample versus maximum rotation
speed, $K$-band luminosity ($M_K$), and $B-K$ color (velocity and
luminosity are tightly correlated via $i_{TF}$). We adjust $q$
for the 3 galaxies in our sample earlier than Sb or later than Scd, as
discussed in DMS-II.
%
%
Our value of $F_{\rm max}^{\rm disk}$ for UGC 463 agrees well within
the errors to the detailed calculation presented in DMS-IV. For the
DMS sample $\langle F_{\rm max}^{\rm disk}\rangle = 0.47 \pm 0.08$
(rms), equivalent to $22^{+8}_{-7}$\% by mass for the disk within
$2.2h_R$, ignoring details of oblateness (the range is marked in
Figure 2c).  In contrast to Herrmann \& Ciardullo (2009), we find
little dependence of $F_{\rm max}^{\rm disk}$ on morphology (Figure
2c).
There is also no significant trend of $F_{\rm max}^{\rm disk}$ with
disk central surface-brightness [$\langle\mu_0(K)\rangle=17.9$ mag
arcsec$^{-2}$, ranging from 16 to 20 mag arcsec$^{-2}$; $18.1$ mag
arcsec$^{-2}$ corresponds to a `Freeman' disk for $B-K=3.5$ mag],
bulge-to-disk ratio ($\langle{\rm B/D}\rangle = 0.1$, ranging from 0 to
0.75), or bar classification. As Figure 2 shows, however, galaxies
with higher rotation speed (see also Kranz et al. 2003), greater
near-infrared luminosity, and redder color are significantly more
maximal.  Adopting a regression model with intrinsic scatter (Akritas
\& Bershady 1996), we find $F_{\rm max}^{\rm disk} = (0.24 \pm 0.08) +
(0.26 \pm 0.08) (V_{\rm max}/200$ km s$^{-1})$ and $F_{\rm max}^{\rm
  disk} = (0.50 \pm 0.07) - (0.06 \pm 0.02)(M_K\!+\!24)$, including
bootstrap errors. The $V_{\rm max}$ trend is consistent with disk
mass-fractions doubling as (total mass)$^{1/2}$.

We have modeled the trend with color assuming galaxies have a constant
mean surface-brightness with color, and estimating changes in
$\Upsilon$ with color based on all SPS models from Bell \& de Jong
(2001), Portinari et al. (2004) and Zibetti et al. (2009). Salient
model variations include treatment of chemical enrichment,
star-formation history, and stellar evolutionary tracks (with varying
amounts of TP-AGB stars). Different mass zeropoints (e.g., IMFs) are
irrelevant here since we normalize the models at $F_{\rm max}^{\rm
  disk} = 0.525$ at $B-K = 3.5$. The curves in Figure 2c show the
median and extrema of these model predictions for {\it relative}
changes in $\Upsilon$ with color. The trends are upper limits since
they assume dynamical mass is equivalent to stellar mass, and discount
contributions from disk gas.

The dependencies of $F_{\rm max}^{\rm disk}$ on color, luminosity and
rotation speed are such that even at extrema (maxima) in each
quantity, the extrapolated value of $F_{\rm max}^{\rm disk} < 0.75$.
For example, we have labeled the $B-K$ colors for different Hubble and
spectral types (Bershady 1995). Our sample spans essentially all of
the color range.  This means that {\it even the earliest-type,
  fastest-rotating, reddest disk should still be submaximal on
  average.} Indeed, the most maximal disk from Herrmann \& Ciardullo
(M94; 2009) has a value of $F_{\rm max}^{\rm disk} \sim
0.7$. Calculations for our sample assume $k=1.5$, i.e., an exponential
disk vertical density distribution. If, for example, disk vertical
density distributions are isothermal, then our estimates of $F_{\rm
  max}^{\rm disk}$ decrease by $\sim$15\%; mass-fractions decrease by
25\%.

These results have significant implications for the IMF, stellar
evolution, cosmological accounting of baryons, and the formation of
galaxy disks. Since a truncated IMF is already required to match
$\Upsilon$ for a maximal disk (Bell \& de Jong 2001), substantially
sub-maximal disks require either uncomfortably top-heavy IMFs, or
validates recent suggestions of a surfeit of luminous, low-mass stars
(e.g., TP-AGB). A quantitative assessment of this surfeit is
forthcoming. Similarly, current accounting of the distribution of
stellar mass in cosmological volumes is based on an $\Upsilon$
calibration that either assumes disks are maximal, or agrees with such
a calibration (e.g., Li \& White 2009, McGaugh et al. 2010). These
stellar mass estimates should be lowered by at least a factor of 3 for
disks, since $F_{\rm max}^{\rm disk}$ represents the total dynamical
mass, not just stellar mass. For example, the stellar mass of the disk
of UGC 463 is only 60\% of the total disk dynamical mass
(DMS-IV). Unless the provocative claim for high $\Upsilon$ in
elliptical cores (van Dokkum \& Conroy 2011) is confirmed and shown to
be more wide-spread, our downward revision of $\Upsilon$ is applicable
to cosmological samples. Finally, galaxy formation models must
reproduce accurate scaling-relations and their scatter (e.g., total
mass to luminosity) in the context of submaximal disks. This
submaximality must correlate with total mass, luminosity and stellar
population, balancing trends of baryon loss from feedback with changes
in angular momentum that, by our reckoning, on average must decrease
threefold from maximum-disk estimates.


Research was supported by grants NSF/AST-9618849, 997078, 0307417,
0607516, 1009491; NSF/OISE-0754437; Spitzer GO-30894; NWO/614.000.807;
LKBF and The Netherlands Research School for Astronomy and
U. Wisconsin Ciriacks Faculty Fellowship in Letters \& Science.

\end{document}